\documentclass[11pt,a4paper]{article}
\usepackage{palatino}
\usepackage{amsfonts,amsmath,amssymb}                     %
\usepackage[a4paper]{geometry}                            %
\usepackage{fancyhdr}                                     %
\usepackage{color}                                        %
\usepackage[pdftex]{hyperref} 
\usepackage{graphicx}
\hypersetup{                                              %
    a4paper,                                              %
    breaklinks                                            %
}                                                         %
{\ \rule{0.5em}{0.5em}}                                   %
\usepackage{subfigure}
\usepackage{tabularx}
\usepackage{booktabs}
\usepackage{quoting}
\usepackage{cite}
\usepackage[all]{nowidow}
\usepackage[font=small]{caption}
\newcolumntype{Y}{>{\centering\arraybackslash}X}

\begin{document}
%
%
\title{\vspace*{-1cm} \bfseries Maximum temporal amplitude and designs of experiments\\ for generation of extreme waves}%
%
%
\author{\begin{tabular}{c}
{\normalsize Marwan{\small$^{1,2}$}, Andonowati{\small $^1$}, and N. Karjanto{\small$^3$}} \\
{\small $^1$Department of Mathematics and Center for Mathematical Modelling and Simulation} \\
{\small Bandung Institute of Technology, Jalan Ganesha 10, Bandung 40132, Indonesia} \\
{\small $^2$Department of Mathematics, Syiah Kuala University} \\
{\small Jalan Teuku Nyak Arief 441, Banda Aceh 23111, Indonesia} \\
{\small $^3$Applied Analysis and Mathematical Physics, Department of Applied Mathematics} \\
{\small University of Twente, PO Box 217, 7500 AE Enschede, The Netherlands}
\end{tabular}
}%
%
\date{}                                                   %
\maketitle                                                %
%
%
\begin{abstract}
\noindent
This paper aims to describe a deterministic generation of extreme waves in a typical towing tank. Such a generation involves an input signal to be provided at the wavemaker in such a way that at a certain position in the wave tank, say at a position of a tested object, a large amplitude wave emerges. For the purpose, we consider a model called a spatial nonlinear Schr\"odinger equation describing the spatial propagation of a slowly varying envelope of a signal. Such a model has an exact solution known as (spatial) Soliton on a Finite Background (SFB) that is a nonlinear extension of Benjamin-Feir instability. This spatial-SFB is characterized by wave focusing leading to almost time-periodic extreme waves that appear in between phase singularities. Although phase singularities and wave focusing have been subject to a number of studies, this spatial-SFB written in the field variables has many interesting properties among which are the existence of many critical values related to the modulation length of the monochromatic signal in the far fields. These properties will be used in choosing parameters for designing experiments on extreme wave generation. In doing so, a quantity called maximum temporal amplitude (MTA) is used. This quantity measures at each location the maximum over time of the wave elevation. For a given modulation length of SFB and desired maximum amplitude at a position in a towing tank, the MTA readily shows the maximum signal that is required at the wavemaker and the amplitude amplification factor of the requested signal. Some examples of such a generation in realistic laboratory variables will be displayed.\\

\setlength{\parindent}{0pt}
\end{abstract}%
%
%
\thispagestyle{fancy}                                     %
%
%
\section{Introduction}

The motivation of this paper stems from the need of hydrodynamic laboratories to generate 'extreme waves', also often called freak or rogue waves that do not break while running downward in the wave tank. In a realistic situation involving large spatial and temporal interval, such a generation is not easy. This is due to the physical limitation of the wavemakers as well as the nonlinear behavior that dominate the deformation of propagating signals from the wavemakers. The dominant nonlinear effects in large wave generations can be seen from the previous theoretical~\cite{brenny99,aan03a}, numerical~\cite{westhuis00,westhuis01} as well as experimental investigation on bi-chromatic waves such as in~\cite{stansberg98}. Depending on the input amplitudes as well as frequencies, large deformation and amplitude increase can be found. The location of the maximum amplitude increase within the wave tank also depends on these parameters.

In a generation of extreme waves in hydrodynamic laboratories, it is desirable that the position for which the extreme waves emerge located at the position of the tested object. A typical question in such a generation is that given the position within the towing tank for the extreme signal to appear, given the amplitude of such a signal within some frequency region, whether a signal can be provided as an input to the wavemakers (with their limitation) such that the requested signal appears in the desired position.

Using an exact solution of a spatial nonlinear Schr\"odinger equation (NLSE) called spatial Soliton on a Finite Background (SFB), it is indeed possible to generate waves of reasonably large within the wave tank of 200~m. This SFB is a nonlinear extension of Benjamin-Feir~\cite{benjamin67} characterized by wave focusing that is almost time periodic. It is recently observed that such an exact solution has numerous interesting properties when it is written in the field variables. Phase singularities; the phenomena of merging and splitting of waves; for which the extreme elevations are sandwiched between them are among these properties. Similar studies on wave focusing and phase singularities have been carried out previously such as in~\cite{karjanto02,aan03b,akhmediev87,ma79,peregrine83,dysthe99}. Here, we are interested in its direct application on the extreme wave generation in a hydrodynamic laboratory.

The content of the paper is as follows. In Section~\ref{mathdescr}, we present a brief overview of the model used in this paper as well as the explicit formula for the spatial SFB. Some interesting features of spatial SFB will also be presented in this section. In Section~\ref{results}, we will show a direct extreme wave generation based on the properties of spatial-SFB and the transformation into laboratory coordinates as well as its interpretation. Examples of realistic time and spatial laboratory scales of this wave generation will be presented here. We present concluding remarks in the final section.

\section{Mathematical description} \label{mathdescr}

Korteweg-de Vries equation is known as an asymptotic model for uni-directional surface gravity waves. If $\eta(x,t)$ denotes the wave elevation, looking for a solution in the form of a harmonic mode centered at a frequency $\omega_0$ modulated by a slowly varying envelope $\psi(\xi,\tau)$, then this complex-valued amplitude $\psi(\xi,\tau)$ satisfies spatial NLSE in the form
\begin{equation*}
\frac{\partial \psi}{\partial \xi} + i \beta \frac{\partial^2 \psi}{\partial \tau^2} + i \gamma |\psi|^2 \psi = 0.
\end{equation*}

Here, the slow variable $\xi = x$ and the shifted time variable $\tau = t - x/\Omega'(k_0)$ are introduced while parameters $\beta$ and $\gamma$ depend on the wavenumber $k_0$ of the monochromatic and the central wave frequency $\omega_0$, where the latter two quantities are related by the linear dispersion relation $\omega_0 = \Omega(k_0) = \sqrt{k_0 \tanh k_0}$, see~\cite{brenny98}.
\begin{figure}[htbp]
\vspace*{-0.5cm}
\begin{center}
\subfigure[]{\includegraphics[width = 0.37\textwidth]{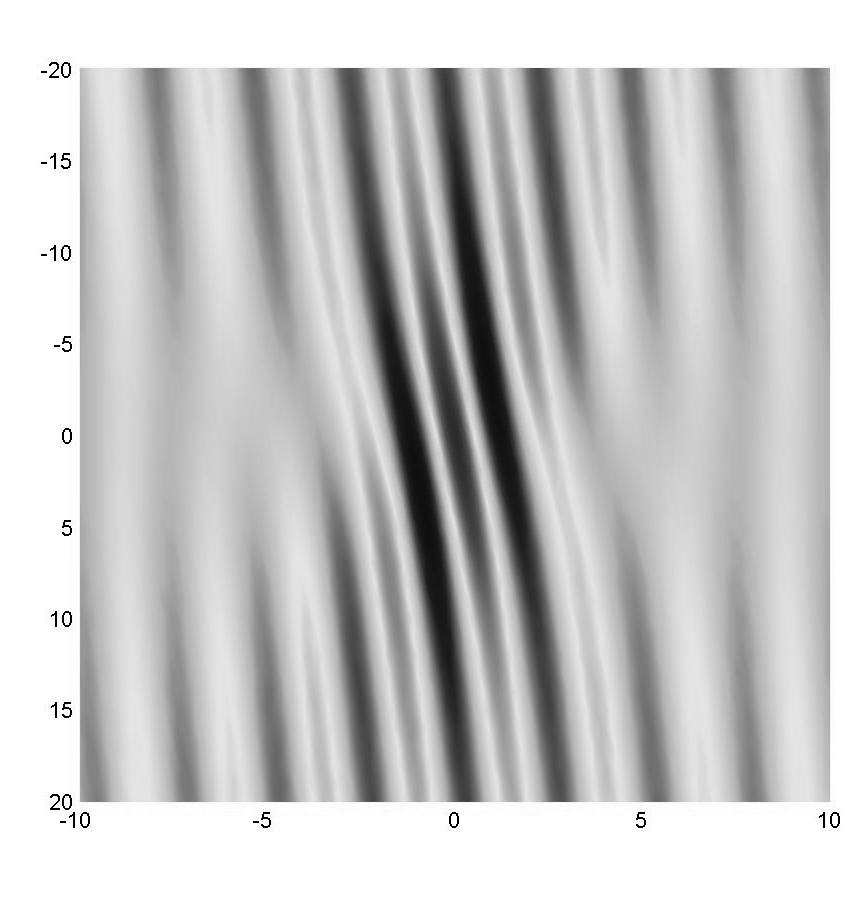}	\label{1a}} \hspace{2cm}
\subfigure[]{\includegraphics[width = 0.40\textwidth]{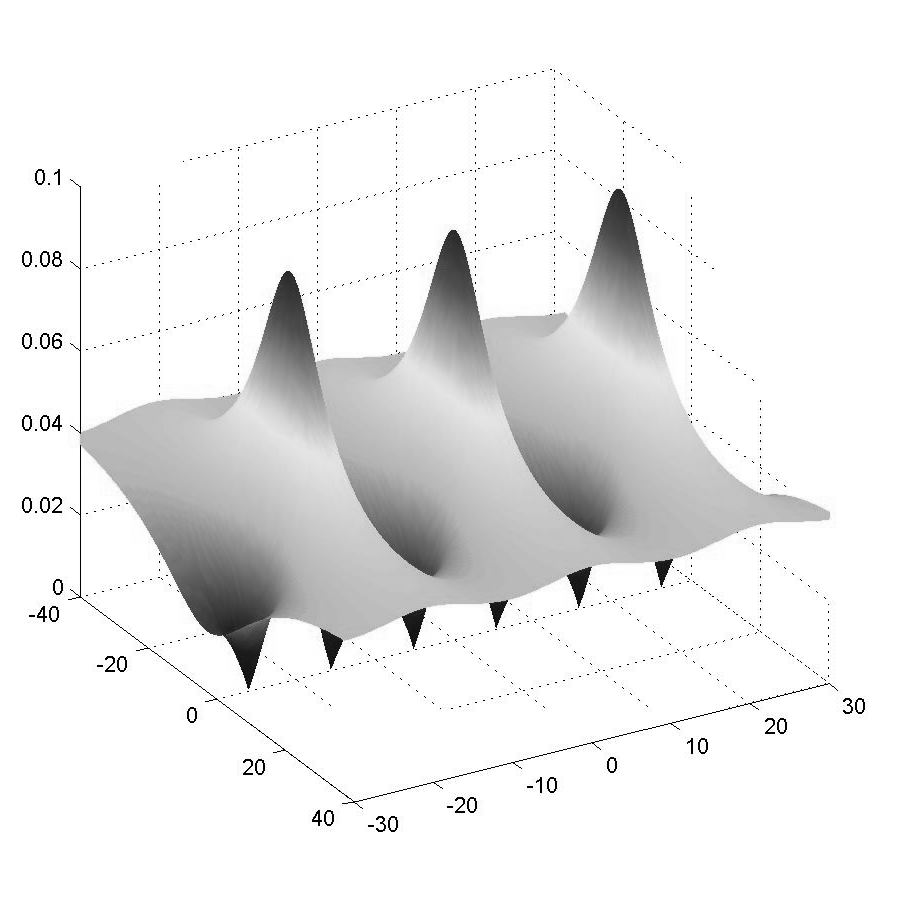} 	\label{1b}}
\end{center}
\vspace*{-0.5cm}
\caption{Contour plot of the wave elevation $\eta(x,t)$ where the extreme wave appears in between the phase singularities signified by merging of two waves into one and splitting one wave into two (left panel) and the corresponding modulus of the complex-valued envelope $|\psi(\xi,\tau)|$ (right panel).} \label{gambar1}
\end{figure}

Different from temporal NLSE that is often used to describe the evolution of ocean waves, spatial NLSE is suitable to describe the propagation of envelope in the signaling problem such as in the problem of wave generation. In~\cite{akhmediev97}, it is shown that the temporal NLSE has an exact solution called SFB. Writing the complex-valued amplitude of such SFB in the form of $\psi(\xi,\tau) = a(\xi,\tau) e^{i\theta(\xi,\tau)}$, then the wave elevation reads $\eta(x,t) = 2 a(\xi,\tau) \cos \phi$ for a real-valued amplitude $a(\xi,\tau)$ and phase $\phi = \theta(\xi,\tau) + k_0 x - \omega_0 t$, with $\omega_0$ and $k_0$ are the frequency and wavenumber of the monochromatic wave, respectively. 

This SFB $\psi(\xi,\tau) = a(\xi,\tau) e^{i\theta(\xi,\tau)}$ has three parameters $(a_0,\nu,\omega_0)$; the real-valued amplitude in the far field $2a_0$, the modulated frequency $\nu$ describing the perturbation on the monochromatic by a long time signal, and the monochromatic wave frequency $\omega_0$. If $\tilde{\nu} = \nu/\nu_\ast$, $\nu_\ast = a_0 \sqrt{\gamma/\beta}$, then $\tilde{\nu}$ lies in the region of Benjamin-Feir instability; $0 < \tilde{\nu} < \sqrt{2}$. The final parameter $\omega_0$ is absorbed in the coefficients of the NLSE equation.

The largest elevation of SFB is at $(\xi,\tau) = (0,0)$ and at a far distance from the position $\xi = 0$, $a(\xi,\tau) \approx a_0$, where the surface wave is monochromatic in time. Some properties of SFB include the amplitude ratio compared to the one of monochromatic in the far field satisfies $1 < a(\xi,\tau)/a_0 < 3$ and in the limiting case $\lim\limits_{\nu \rightarrow 0} a(\xi,\tau)/a_0 = 3$ giving the largest possible amplification factor of three from this monochromatic background.
\begin{figure}[htbp]
\vspace*{-0.4cm}
\begin{center}
\includegraphics[width = 0.8\textwidth, height= 5cm]{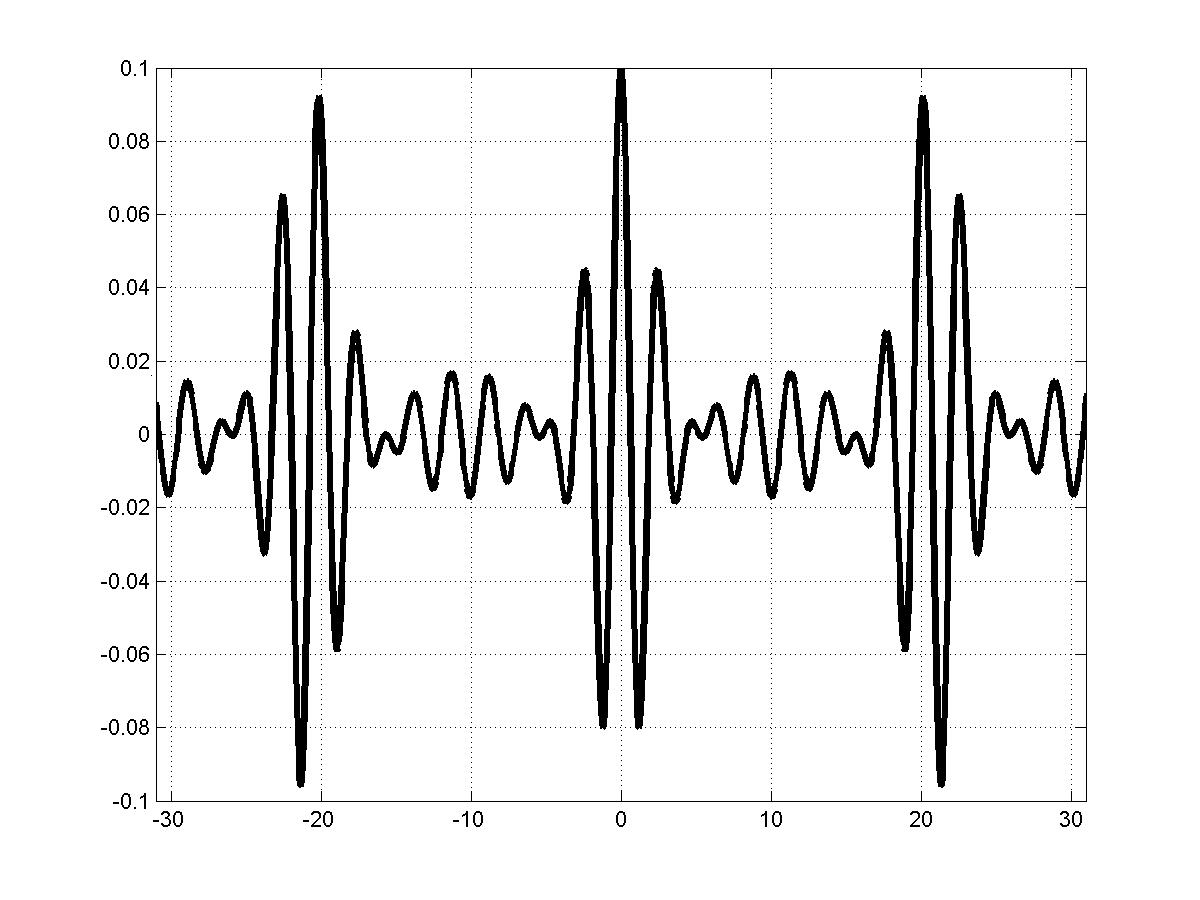}
\end{center}
\vspace*{-0.5cm}
\caption{A wave signal at the extreme position $\xi = 0$ for $\omega_0 = 2.5$, $a_0 = 0.0207$ and $\nu = 0.3014$.} \label{gambar2}
\end{figure}

It has been known that these extreme elevations appear in between two phase singularities; a phenomenon when two waves are merging into one wave or one wave is splitting into two. The phenomenon can be seen in a close-up contour plot of the elevation in Figure~\ref{1a}. The envelope corresponding to this spatial SFB is shown in Figure~\ref{1b} where the locations (or instants) of the phase singularities, also known as `wavefront dislocations'~\cite{nye97}, are clearly visible; they are the instants when the envelope touches the $xt$-plane. 

In Figure~\ref{gambar2}, we display a time signal with an SFB envelope at $\xi = 0$ for the three parameter values $\omega_0 = 2.5$, $a_0 = 0.0207$ and $\nu = 0.3014$ and so $T = 2\pi/\nu = 28.8458$. We observe that the extreme elevation is almost repeated in time with periodicity of $T$.

To give an easier interpretation related to the problem of wave generation later, in what follows we require that at $(\xi,\tau) = (0,0)$, $\psi(0,0) = m_0$. Let suppose that the frequency of the monochromatic wave $\omega_0$ is given, which is the center frequency of the wave group being generated. If the amplification factor $\alpha$ is defined as a quotient between the elevation at the extreme position and in the far field, that is $\alpha = m_0/a_0$, then the three parameters of SFB $(a_0,\nu,\omega_0)$ can be recovered from $a_0 = m_0/\alpha$ and $\nu = \nu(\omega_0,\alpha) = a_0 \sqrt{\gamma(\omega_0)/\beta(\omega_0)} \sqrt{2 - (\alpha - 1)^2/2}$.

\section{Results} 	\label{results}

In what follows, we use the laboratory coordinates for the case of a wave tank with depth $H$ of 5~m. The wave tank at MARIN (Maritime Research Institute Netherlands, Wageningen) is taken as reference tank; see~\cite{westhuis01}. The length of that tank is 200~m, and thus the extreme waves and the amplitude amplification should appear within a distance of 200~m from the wavemaker. It is our intention to be able to compare the results here with numerical computation obtained from a numerical wave tank as well as measurement in the near future. Although here we do not describe precisely on the technical restriction of wavemaker, our aim is that the required signal can be fed easily to the wavemakers.

Let $x_\text{wg} = 0$ be the location of the wavemaker and $x_\text{ship}$ be the location in the tank such that the signal generated at the wavemaker achieves its extreme. If $\eta_\text{lab}(x_\text{lab},t_\text{lab})$ is the elevation of a signal with SFB envelope in the laboratory variables at $x_\text{lab}$, let us define at a given position, the maximum temporal amplitude (MTA) in the form
\begin{equation*}
M(x_\text{lab}) = \max_{t_\text{lab}} \; \eta_\text{lab} (x_\text{lab}, t_\text{lab})
\end{equation*}
and $x_\text{ship}$ corresponding to $\xi = 0$ in NLSE spatial variable and so the translation from the laboratory variables to NLSE variables are as follows $\xi = x - x_\text{ship}$ and $\tau = t - (x - x_\text{ship})/\Omega'(k_0)$, where $\omega_0 = \Omega(k_0) = \sqrt{k_0 \tanh k_0}$ and $\omega_0 = \omega_\text{lab}/\sqrt{g/H}$ for the laboratory frequency $\omega_\text{lab}$ and gravitational acceleration $g$.
\begin{table}[h]
\begin{center}
\begin{tabularx}{\textwidth}{@{}c*{6}{Y}c@{}}
\toprule
Example & $\alpha$ & $\tilde{\nu}$  & $2a_0^\text{lab}$ & $\nu_\text{lab}$  & $T_\text{lab}$ & $M(0)$ & $\alpha_\text{actual}$ \\ \hline
1       & 2.0000   & $\sqrt{3/2}$ & 0.2500            & 0.6238            & 10.0723        & 0.2707 & 1.8468                 \\
2       & 2.2642   & 1.0959       & 0.2208            & 0.4930            & 12.7439        & 0.2481 & 2.0151                 \\ 
3       & 2.4142   & 1       	  & 0.2071	          & 0.4220	  	      & 14.8906        & 0.2391 & 2.0916   		         \\ 
4	    & 2.7321   & $1/\sqrt{2}$ & 0.1830	          & 0.2636			  & 23.8337        & 0.2276 & 2.1967 			     \\ 
5	    & 2.9412   & 0.3405  	  & 0.1700	          & 0.1179			  & 53.2941        & 0.2247 & 2.2253 			     \\ 
\bottomrule
\end{tabularx}
\end{center}
\vspace*{-0.5cm}
\caption{Samples of extreme wave generation based on SFB. The position of the ship ($x_\text{ship}$) is taken to be 125~m from the wavemaker and the MTA at the ship position is $M(x_\text{ship}) = 0.5$~m.} \label{table} 
\end{table}

The amplification factor $\alpha = M(x_\text{ship})/2a_0^\text{lab}$ where $2a_0^\text{lab}$ is the amplitude in the laboratory variable of the monochromatic wave in the far field. The actual amplification factor is given by $\alpha_\text{actual} = M(x_\text{ship})/M(0)$. In what follow we list a number of examples of extreme waves generated based on SFB. The position of the ship is taken to be 125~m from the wavemaker while the MTA of the signal at the ship position is $M(x_\text{ship}) = 0.5$~m and $\omega_\text{lab} = 3.5$.
\begin{figure}[htbp]
\vspace*{-0.5cm}
\begin{center}
\subfigure[]{\includegraphics[width = 0.4\textwidth]{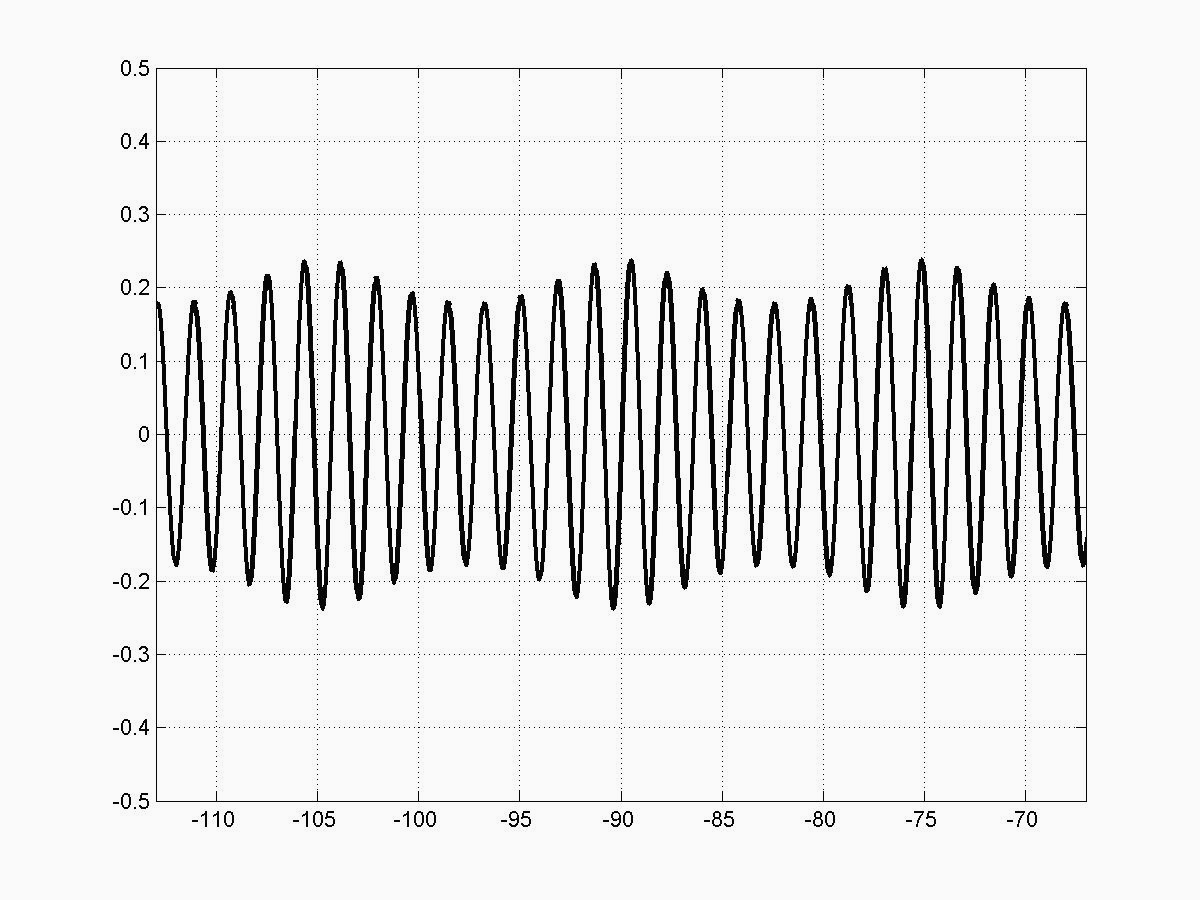}	\label{3a}} \hspace{2cm}
\subfigure[]{\includegraphics[width = 0.4\textwidth]{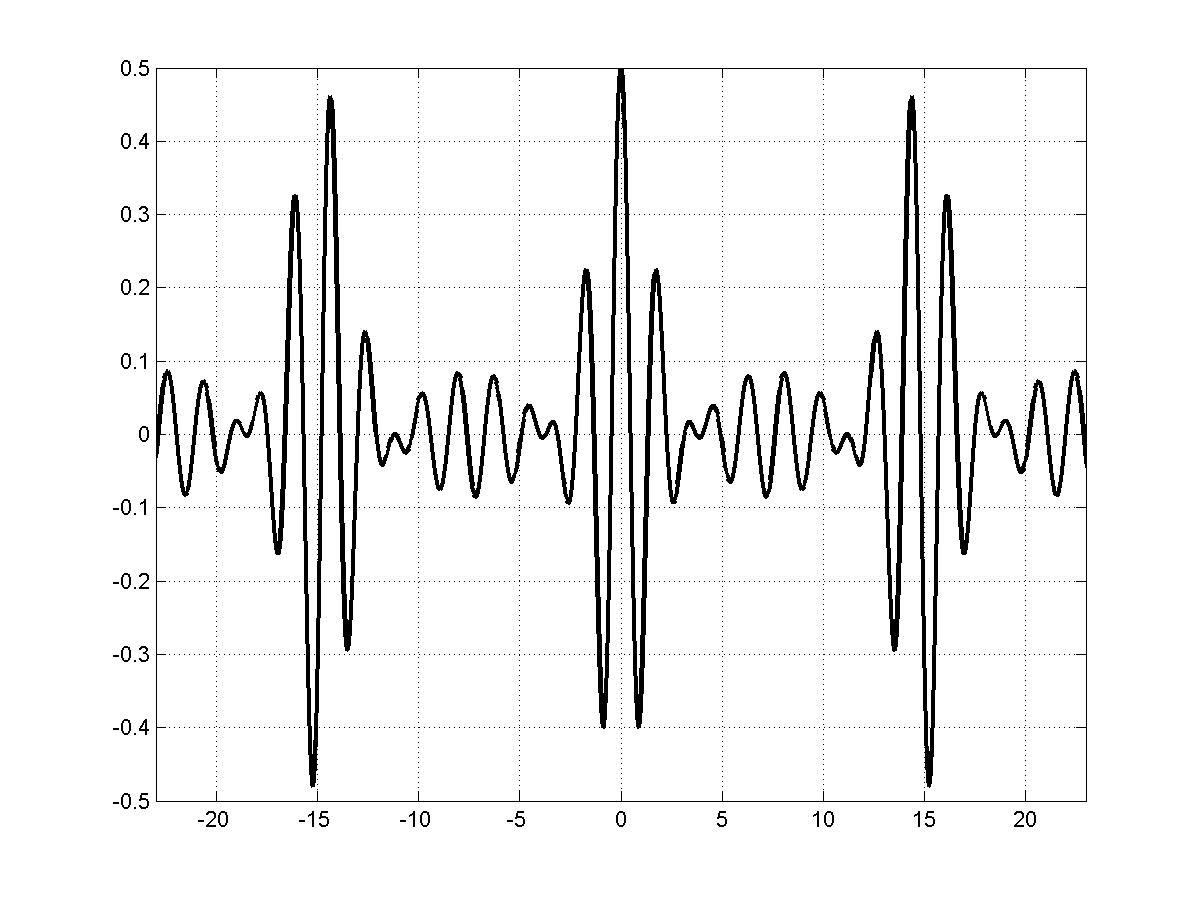} 	\label{3b}}
\end{center}
\vspace*{-0.5cm}
\caption{Wave signals for $\alpha = 2.4142$ at the wavemaker (left panel) and at 125~m from the wavemaker (right panel).} \label{gambar3}
\end{figure}

\begin{figure}[htbp]
\vspace*{-0.5cm}
\begin{center}
\subfigure[]{\includegraphics[width = 0.4\textwidth]{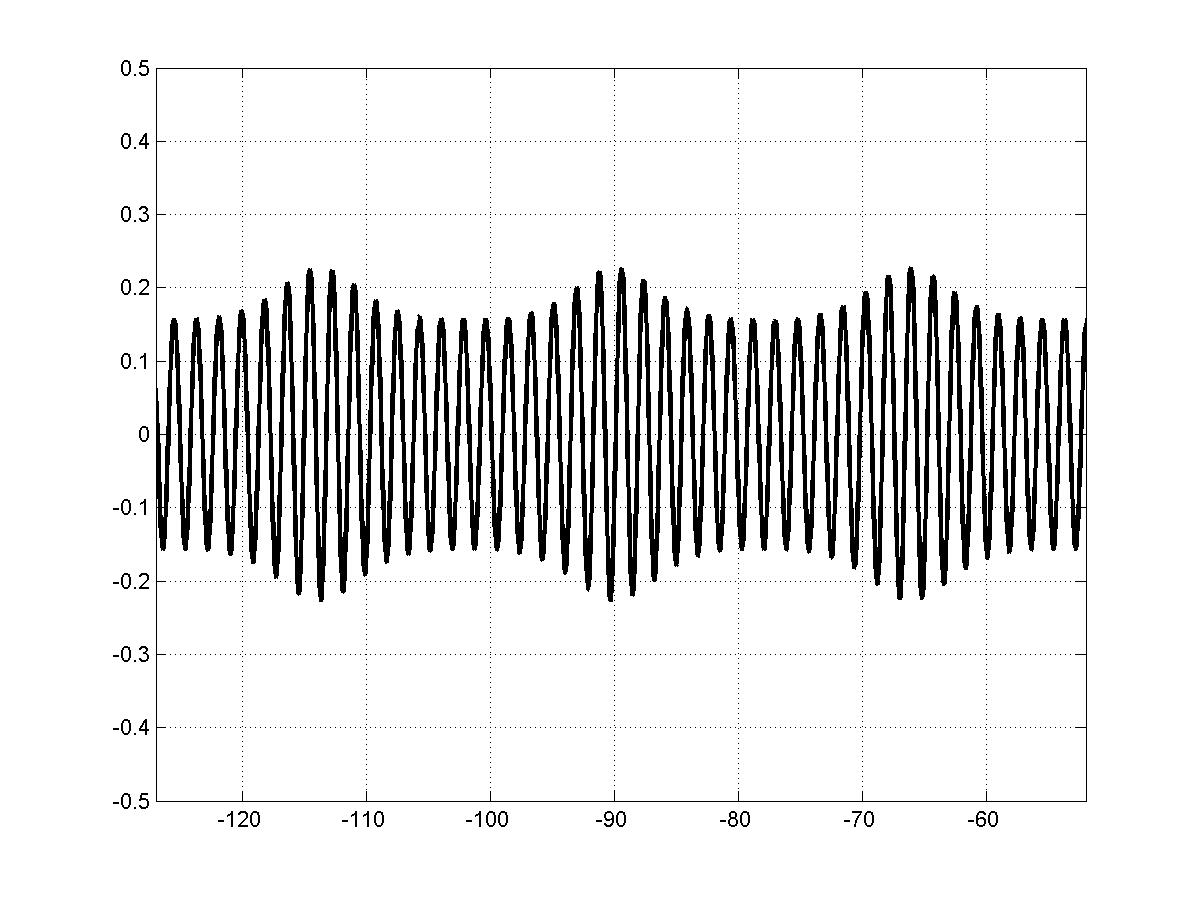}	\label{4a}} \hspace{2cm}
\subfigure[]{\includegraphics[width = 0.4\textwidth]{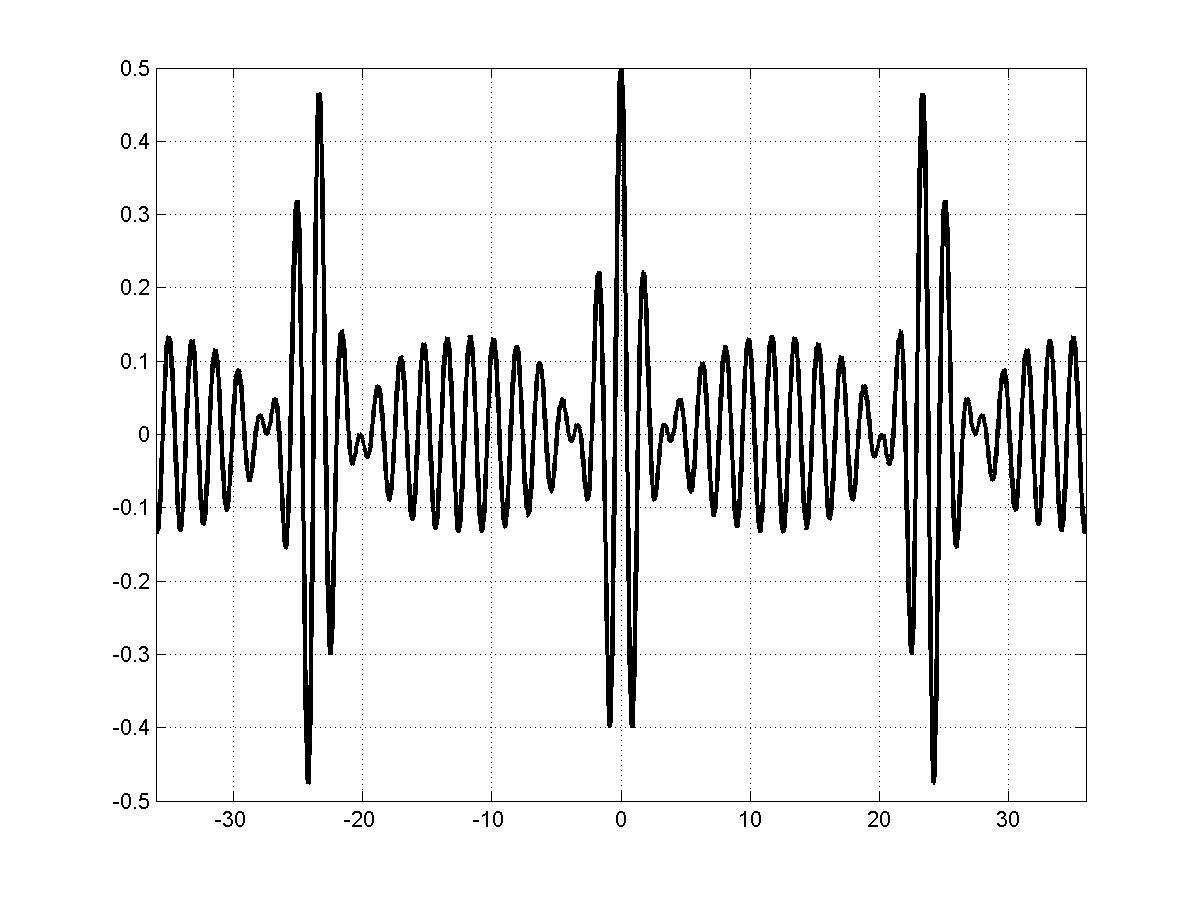} 	\label{4b}}
\end{center}
\vspace*{-0.5cm}
\caption{Wave signals for $\alpha = 2.7321$ at the wavemaker (left panel) and at 125~m from the wavemaker (right panel).} \label{gambar4}
\end{figure}

\begin{figure}[htbp]
\vspace*{-0.4cm}
\begin{center}
\includegraphics[width = 0.8\textwidth, height = 8cm]{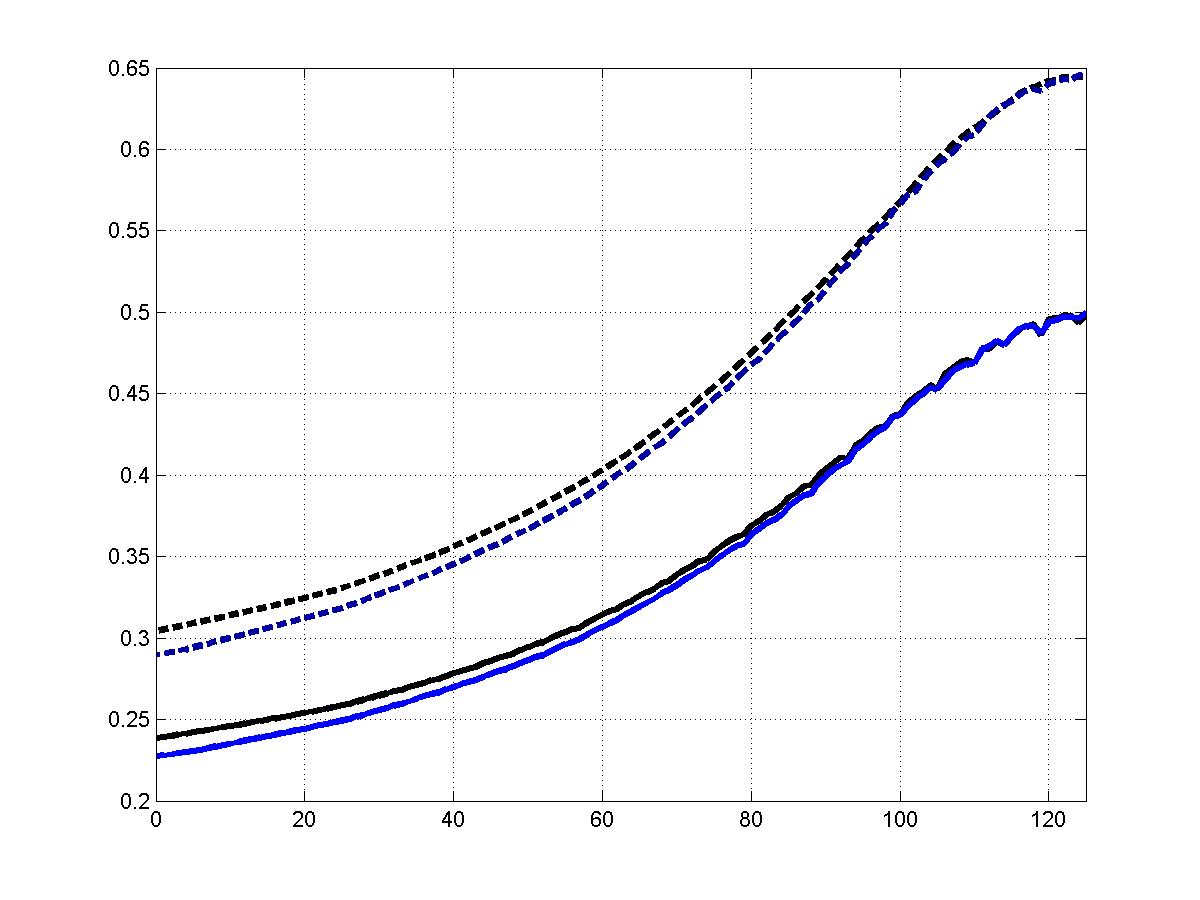}
\end{center}
\vspace*{-0.5cm}
\caption{Maximum temporal amplitude (solid lines) and maximum temporal steepness (dashed lines) for $\alpha = 2.4142$ (black, upper part) and $\alpha = 2.7321$ (blue, lower part) corresponding to $\tilde{\nu} = 1$ and $\tilde{\nu} = \sqrt{2}$, respectively.} \label{gambar5}
\end{figure}

Further, we plot wave signals at the wavemaker and at the ship position 125~m away from the wavemaker for $\alpha = 2.4142$ and $\alpha = 2.7321$, corresponding to Examples~3 and~4 in Table~\ref{table}. See Figures~\ref{gambar3} and~\ref{gambar4}, respectively. The MTA $M(x_\text{lab})$, as well as the corresponding maximum temporal steepness, defined as the product of wave elevation and local steepness, $\eta_\text{lab} \, k_\text{lab}$, for both examples are displayed in Figure~\ref{gambar5}.

\section{Concluding remarks} \label{conclusion}

We have considered a model called spatial NLSE that is suitable to describe the propagation of slowly varying envelopes in signaling problems such as in the wave generation often performed in a hydrodynamic laboratory. This NLSE has an exact solution called spatial SFB with interesting properties related to wave focusing and phase singularity. Written in the field variables, it is interesting to see that the extreme waves appear in between the phase singularities signified by merging of two waves into one and splitting one wave into two.

We have used properties of SFB in the generation of extreme waves in realistic laboratory coordinates. Examples of realistic time and spatial laboratory scales of this wave generation have been presented. The results show that the generation of such extreme waves is possible to be conducted in hydrodynamic laboratories. Future research will focus on both numerical and experimental evidence to validate the theoretical prediction.  

\subsection*{Acknowledgement}
{\small The authors are very grateful to Prof. E. (Brenny) van Groesen (University of Twente and LabMath Indonesia) for insight and discussion throughout the execution of this research. They would like to thank Dr. Gert Klopman (University of Twente and Witteveen$+$Bos, The Netherlands) for many useful hints on the wave generation. This research is conducted party at Bandung Institute of Technology, Indonesia and partly at the University of Twente, The Netherlands and is funded by the Indonesian International Research Linkage Programme (RUTI 2002/2005) in a framework of international cooperation with the University of Twente under the project `Prediction and Generation of Deterministic  Extreme  Waves in Hydrodynamic Laboratories' (TWI.5374) of the Netherlands Organization of Scientific Research NWO, subdivision Applied Sciences STW. \par}

\end{document}